\documentclass[%
 reprint,
nofootinbib,
 amsmath,amssymb,
 aps,
 showkeys,
]{revtex4-1}

\usepackage{graphicx}
\usepackage{dcolumn}
\usepackage{bm}

\begin{document}

\title{ A Classical  $\pi$ Machine and Grover's Algorithm}
\author{Jiang Liu}
\email{jiangliu@alumni.cmu.edu}

\date[]{January 5, 2020}

\begin{abstract}
This paper studies a well-known $\pi$ machine illustrated by Fig.~(\ref{fig:ClassicalPiMachine}). It is shown that the $\pi$ machine can compute digits of $\pi$ if the ratio of block weights, $m_2/m_1$, satisfies certain conditions, and that dynamics of the $\pi$ machine is identical to that of Grover's algorithm \cite{Grover} in quantum computing.

\end{abstract}

                             
\maketitle


\section{Introduction}

\begin{figure}[h]
	\centering
	\includegraphics[scale=0.6]{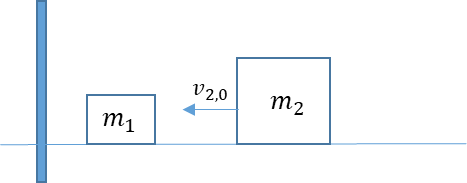}
	\caption{	\label{fig:ClassicalPiMachine}
		Initial state of two sliding block elastic collision $\pi$ machine.
	}
\end{figure}

Computation of $\pi$ has never ceased to delight us\footnote{Current world record of $\pi$ digits is 31.4 trillion by Emma Haruka Iwao using Chudnovsky algorithm \cite{Chudnovsky1988} on Google Cloud \cite{Iwao2019,Kleinman2019,Shaban2019}}. 
One intereting method\cite{Illner2013, Antonick2014,Aretxabaleta2017,Fraser2019} is shown in Fig.~(\ref{fig:ClassicalPiMachine}), 
where digits of $\pi$ is computed by counting the number of elastic collisions between two sliding blocks of masses $m_1$ and $m_2$,  and collisions  between block $m_1$ and  a wall.  This paper is to take a fresh look at this simple machine.

 The rest of this paper is organized as follows. Section II shows that  calculability of $\pi$ digits depends on $m_2/m_1$. Section III shows that dynamics of the $\pi$ machine is the same as that of  Grover's algorithm \cite{Grover}, and that Grover's quantum computing \cite{NC2010} can be visualized classically by the $\pi$ machine. Our conclusion is summarized in Section V.

\section{The $\pi$ Machine}

\begin{figure}[h]
	\centering
	\includegraphics[scale=0.6]{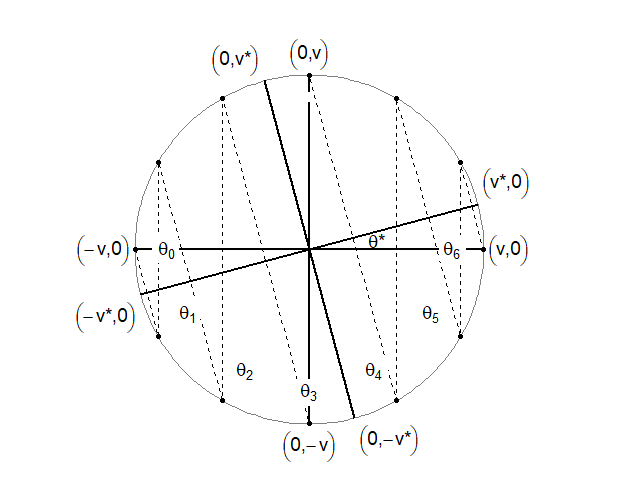}
	\caption{	\label{fig:PiMachine}
		Geometric representation of a $\pi$ machine's dynamics whose $\theta^*=\pi/12$. 
		Doted lines parallel to  $(0,v)-(0,-v)$ axis, represent $m_1$ bouncing off the wall. 
		Doted lines parallel to  $(0,v^*)-(0,-v^*)$ axis represent $m_1-m_2$ collision. 
		$\theta_t$ is the polar angle of $\bm{v}_t$ after each $m_1-m_2$ collision.
	}
\end{figure}

Consider a weighted velocity space shown by Fig.~(\ref{fig:PiMachine}), where
\begin{equation}
\bm{v}_t \equiv 
\begin{pmatrix}
&\sqrt{m_2}v_{2,t}\\
&\sqrt{m_1}v_{1,t}
\end{pmatrix}
\label{eqn:BasisVector}
\end{equation}
and $v_{1,t}$ and $v_{2,t}$ are the velocities of $m_1$ and $m_2$ at $t$.
Conservation of kinetic energy 
requires that $\bm{v}_t$ be on a circle of radius $\vert \bm{v}_t \vert =  \sqrt{m_2}v_{2,0}$.
Conservation of momentum  implies
\begin{equation}
m_{2}v_{2,t+1}+m_{1}v_{1,t+1} = m_2v_{2,t}-m_1v_{1,t}, 
\label{eqn:MConservation}
\end{equation}
where minus sign of $m_1v_{1,t}$ rises from $m_1$ bouncing off the wall. In polar coordinates, where   $\sqrt{m_2}v_{2,t}= \vert \bm{v}_t\vert \cos\theta_t$ and $\sqrt{m_1}v_{1,t}= \vert \bm{v}_t\vert \sin\theta_t$, Eq.~(\ref{eqn:MConservation}) can be simplified to
\begin{equation}
\cos(\theta_{t+1} - \theta^*) = \cos(\theta_{t}+\theta^*),
\label{eqn:theta_t}
\end{equation}
 where
\begin{equation}
\sin\theta^* = \sqrt{\frac{m_1}{m_1+m_2}}.
\label{eqn:theta_0}
\end{equation}

Solution to Eq.~(\ref{eqn:theta_t}) with initial condition $\theta_0 = \pi$ is
\begin{equation}
\theta_t  = 2t\theta^*+\pi.
\label{eqn:classical}
\end{equation}
Collision  stops at $T$ when $v_{1,T}\to 0$, i.e., $\theta_T\to 2\pi$.
Total number of collisions between $t=0$ and $t=T$ is $2T$ and from Eq.~(\ref{eqn:classical})
\begin{equation}
2T = \left\lfloor \frac{\pi}{\theta^*}\right\rfloor .
\label{eqn:ClassicalResult}
\end{equation}

Clearly, number of collisions prints digits of $\pi$ if $\theta^{*} = 10^{-n}$, where $n$ is a positive integer. This happens when $m_2/m_1 = 10^{2n}\gg 1$ (see Eq.~(\ref{eqn:theta_0})). This is the necessary condition for the calculability of the $\pi$ machine.  Otherwise, the $\pi$ machine cannot produce digits of $\pi$.

\section{The $\pi$ Machine and Grover's Algorithm}

That dynamics of the $\pi$ machine and Grover's algorithm  \cite{Grover} are identical can be viewed most conveniently from  Hamiltonian formulation,
in which collisions are described by evolution of states of bases $\bm{v}$ and $\bm{v^*}$.

In $\bm{v}$ basis, $\bm{v}_t$ is an eigenvector of Hamiltonian for  $m_1$ bouncing off the wall.  Geometrically, $\bm{v}_t$ reflects over the $(v,0)$  axis of Fig.~(\ref{fig:PiMachine}), i.e.,
\begin{equation}
\bm{v}_t 
\rightarrow
\begin{pmatrix}
1 & 0\\
0 & -1
\end{pmatrix}
\bm{v}_t
\label{eqn:ComputationalBasis}.
\end{equation}
$\bm{v}$ basis is referred to as computational basis in the literature. It is called computational due to interactions between internal system (blocks of the $\pi$ machine) and environment(the wall).

$\bm{v^*}$ basis is related to $\bm{v}$ basis by a rotation (see Fig.~(\ref{fig:PiMachine})) with $\theta^*$ given by Eq.~(\ref{eqn:theta_0})
\begin{equation}
\bm{v^*}_t =  U_{\theta^*}\bm{v}_t,
\label{eqn:BasisChange}
\end{equation}
where
\begin{equation}
U_{\theta^*}=
\begin{pmatrix}
\cos\theta^*  &  \sin\theta^*\\
-\sin\theta^* &  \cos\theta^*
\end{pmatrix}.
\label{eqn:Rotation}
\end{equation}

In $\bm{v^*}_t$  basis, $\bm{v^*}_t$ is an eigenvector of Hamiltonian for  block collision. Geometrically,  $\bm{v^*}_t$ reflects over the $(v^*,0)$  axis of Fig.~(\ref{fig:PiMachine}), i.e.,
\begin{equation}
\bm{v^*}_{t + 1}
=
\begin{pmatrix}
1 & 0\\
0 & -1
\end{pmatrix}
\bm{v^*}_t
\label{eqn:CanonicalBasis}.
\end{equation}
$\bm{v^*}$ basis is referred to as canonical basis in the literature. Canonical basis does not involve environment.

It then follows from Eqs.~(\ref{eqn:ComputationalBasis}) to (\ref{eqn:CanonicalBasis})
that dynamics of the $\pi$ machine in $\bm{v}$ basis (computational basis) is
\begin{equation}
\bm{v}_{t+1} = G_{\theta^*}
\bm{v}_t,
\label{eqn:Evolution}
\end{equation}
where 
\begin{equation}
G_{\theta^*}
=
U_{\theta^*}^{\dagger}
\begin{pmatrix}
1 & 0\\
0 & -1
\end{pmatrix}
U_{\theta^*}
\begin{pmatrix}
1 & 0\\
0 & -1
\end{pmatrix}.
\label{eqn:EvolutionMatrix}
\end{equation}
Upper component of Eq.~(\ref{eqn:Evolution}) is Eq.~(\ref{eqn:theta_t}) discussed in the previous section.
Lower component  of Eq.~(\ref{eqn:Evolution}) is the conjugate of Eq.~(\ref{eqn:theta_t}),
$\sin(\theta_{t+1}-\theta^*) = \sin(\theta_t+\theta^*)$. Both lead to the same result.

Apart from an overall phase of no physical significance, dynamics described by Eqs.~(\ref{eqn:Evolution}) and (\ref{eqn:EvolutionMatrix}) is the same as that of Grover's diffusion operator \cite{Grover,NC2010}. 

\begin{figure}[h]
	\centering
	\includegraphics[scale=0.8]{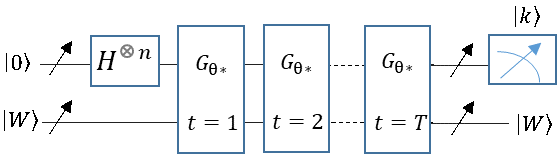}
	\caption{\label{fig:QPiMachine}
		Quantum $\pi$ machine simulator (Grover's circuit). $\vert W\rangle$ is Grover's working qubit representing the wall. $\vert 0\rangle$ is the initial $n$-qubit state in canonical basis representing  block state $(-\bm{v},0)$.   $\vert k\rangle$ is the search target state in computational basis representing block state $\sqrt{m_1}v_{1,0}$.
	}
\end{figure}

As a result, processes described by Fig.~(\ref{fig:PiMachine}) for the $\pi$ machine can be interpolated into Grover's quantum circuits shown by Fig.~(\ref{fig:QPiMachine}). This interpolation allows us to visualize Grover's quantum computing by the $\pi$ machine.

Working qubits $\vert W\rangle$ in Grover's algorithm can be visualized as the wall, searching target $\vert k\rangle$  as
$\sqrt{m_1}v_{1,0}$ and superposition of the rest of qubits as $\sqrt{m_2}v_{2,0}$; Total number of qubits $n$ is determined by 
$n = \lceil\log_2( 1+ m_2/m_1) \rceil$.  

Grover's diffusion operator $G_{\theta^*}$ can be visualized as $m_1$ bouncing off the wall in computational basis, which plays the same role as Grover's Oracle, i.e., $\vert k\rangle \to - \vert k\rangle$, followed by $m_1$ and $m_2$ collision in canonical basis, which plays the same role as phase shift, i.e., $\vert 0\rangle \to -\vert 0\rangle$.
Number of diffusion operators can be visualized as number of collisions.

Most remarkably, basis mixing Eq.~(\ref{eqn:Rotation}), which is generated in Grover's algorithm by Hadamard gate rotating qubit along polar axis of Bloch Sphere\cite{Bloch46}, is a consequence of conservation of kinetic energy and momentum of the $\pi$ machine.

There is one difference between the $\pi$ machine and Grover's search. In the $\pi$ machine, initial state can be prepared and final state can be measured in computational basis alone. In Grover's search, however, initial state $\vert 0\rangle$ is prepared in canonical basis and final state is measured in computational basis. Probability of finding $\vert k\rangle$ of computational basis, which is equivalent to measuring $\sqrt{m_1}v_{1,t}$,  from initial state $\vert 0\rangle$ of canonical basis  is
\begin{equation}
Pr(\langle k \vert 0 \rangle)_t = \sin^2\theta_t,
\end{equation}
where $\theta_t = 2t\theta^* + \theta^*$  (mod $\pi$), which has 
an extra $\theta^*$ when compared to Eq.~(\ref{eqn:classical}).

Apart from the difference of initial state, dynamics of the $\pi$ machine and Grover's algorithm are identical. Computational complexity of these two processes is the same. In order words, the classical $\pi$ machine is effectively simulating a type of Grover's quantum computing.


\section{Conclusion}

Results of this paper can be summarized as follows: digits of $\pi$ can be computed by the $\pi$ machine only in limited cases. Dynamics of the $\pi$ machine is the same as that of Grover's algorithm in all cases. Grover's quantum computing may be classically visualized by the $\pi$ machine.

After the completion of this work, I noticed from Quanta Magazine \cite{GSanderson2020} a very interesting recent work of Adam Brown \cite{ABrown2019} on the same subject. Although our approaches are somewhat different, we independently reached the same conclusion.

\begin{acknowledgments}
	I would like to thank Ken Li for valuable discussions. He verified numerically the validity of Eqs.~(\ref{eqn:theta_0}) and (\ref{eqn:ClassicalResult}).
	
	Any views expressed are mine as an individual and not as a representative speaking for or on behalf of my employer, nor do they represent my employer's positions, strategies or opinions.
	
	This research did not receive any specific grant from funding agencies in the public, commercial, or not-for-profit sectors.
	
	Declarations of interest: None.
\end{acknowledgments}

\nocite{*}

\bibliography{Pi2}

\end{document}